\documentclass[11pt,a4paper]{article}
\usepackage{amsfonts}
\usepackage{amsmath,amssymb}
\usepackage{epsfig,graphicx}

\input{tcilatex}

\begin{document}

\begin{center}
{\LARGE \ Graviton as a Goldstone boson: } \smallskip\smallskip

{\LARGE Nonlinear Sigma Model for Tensor Field Gravity }

\bigskip \bigskip

\textbf{J.L.~Chkareuli}$^{1,2}$, \textbf{J.G. Jejelava}$^{1,2}$\textbf{\ and
G. Tatishvili}$^{2}$ \bigskip

$^{1}$\textit{E. Andronikashvili} \textit{Institute of Physics, 0177
Tbilisi, Georgia\ }

$^{2}$\textit{Center for Elementary Particle Physics, ITP,} \textit{Ilia
State University, 0162 Tbilisi, Georgia\ \vspace{0pt}\\[0pt]
}

\bigskip

\bigskip

\bigskip

\bigskip

\bigskip

\bigskip

\bigskip

\bigskip

\textbf{Abstract}

\bigskip
\end{center}

Spontaneous Lorentz invariance violation (SLIV) realized through a nonlinear
tensor field constraint $H_{\mu \nu }^{2}=\pm M^{2}$ ($M$ is the proposed
scale for Lorentz violation) is considered in tensor field gravity theory,
which mimics linearized general relativity in Minkowski space-time. We show
that such a SLIV pattern, due to which the true vacuum in the theory is
chosen, induces massless tensor Goldstone modes some of which can naturally
be associated with the physical graviton. When expressed in terms of the
pure Goldstone modes, this theory looks essentially nonlinear and contains a
variety of Lorentz and $CPT$ violating couplings. Nonetheless, all SLIV
effects turn out to be strictly cancelled in all the lowest order processes
considered, provided that the tensor field gravity theory is properly
extended to general relativity (GR). So, as we generally argue, the
measurable effects of SLIV, induced by elementary vector or tensor fields,
are related to the accompanying gauge symmetry breaking rather than to
spontaneous Lorentz violation. The latter appears by itself to be physically
unobservable, only resulting in a non-covariant gauge choice in an otherwise
gauge invariant and Lorentz invariant theory. However, while Goldstonic
vector and tensor field theories with exact local invariance are physically
indistinguishable from conventional gauge theories, there might appear some
principal distinctions if this local symmetry were slightly broken at very
small distances controlled by quantum gravity in an explicit, rather than
spontaneous, way that could eventually allow one to differentiate between
them observationally. 

%%%%%%%%%%%%%%%%%%%%%%%%%%%%%%%%%%%%%%%%%%%%%%%%%%%%%%%

\thispagestyle{empty}\newpage

\section{Introduction}

It is no doubt an extremely challenging idea that spontaneous Lorentz
invariance violation (SLIV) could provide a dynamical approach to quantum
electrodynamics \cite{bjorken}, gravity \cite{ohan} and Yang-Mills theories 
\cite{eg} with photon, graviton and non-Abelian gauge fields appearing as
massless Nambu-Goldstone (NG) bosons \cite{NJL} (for some later developments
see \cite{cfn,jb,kraus,jen,bluhm}). This idea has recently gained new
impetus in the gravity sector - as for composite gravitons \cite{kan}, so in
the case when gravitons are identified with the NG modes of the symmetric
two-index tensor field in a theory preserving a diffeomorphism (diff)
invariance, apart from some non-invariant potential inducing spontaneous
Lorentz violation \cite{kos,car}.

We consider here an alternative approach which has had a long history,
dating back to the model of Nambu \cite{nambu} for QED in the framework of
nonlinearly realized Lorentz symmetry for the underlying vector field. This
may indeed appear through the "length-fixing" vector field constraint

\begin{equation}
A_{\mu }^{2}=n^{2}M^{2}\text{ , \ \ }n^{2}\equiv n_{\nu }n^{\nu }=\pm 1
\label{const}
\end{equation}%
(where $n_{\nu }$ is a properly oriented unit Lorentz vector, while $M$ is
the proposed scale for Lorentz violation) much as it works in the nonlinear $%
\sigma $-model \cite{GL} for pions, $\sigma ^{2}+\pi ^{2}=f_{\pi }^{2}$,
where $f_{\pi }$ is the pion decay constant. Note that a correspondence with
the nonlinear $\sigma $ model for pions may appear rather suggestive in view
of the fact that pions are the only presently known Goldstone particles
whose theory, chiral dynamics\cite{GL}, is given by the nonlinearly realized
chiral $SU(2)\times SU(2)$ symmetry rather than by an ordinary linear $%
\sigma $ model \footnote{%
Another motivation for the constraint (\ref{const}) might be an attempt to
avoid an infinite self-energy for the electron in classical electrodynamics,
as was originally suggested by Dirac \cite{dir} (and extended later to
various vector field theories \cite{ven}) in terms of the Lagrange
multiplier term, $\frac{1}{2}\lambda (A_{\mu }^{2}-M^{2})$, due to which the
constraint (1) appears as an equation of motion for the auxiliary field $%
\lambda (x)$. Recently, there was also discussed in the literature a special
quadratic Lagrange multiplier potential \cite{kkk}, $\frac{1}{4}\lambda
\left( A_{\mu }^{2}-M^{2}\right) ^{2},$ leading to the same constraint (\ref%
{const}) after varying the action, while the auxiliary $\lambda $ field
completely decouples from the vector field dynamics rather than acting as a
source of some extra current density, as it does in the Dirac model.
Formally, numbers of independent degrees of freedom in these models appear
different from those in the Nambu model \cite{nambu}, where the SLIV
constraint is proposed to be substituted into the action prior to varying of
the action. However, in their ghost-free and stability (positive
Hamiltonian) phase space areas \cite{kkk} both of them are physically
equivalent to the Nambu model with the properly chosen initial condition.}.
The constraint (\ref{const}) means in essence that the vector field $A_{\mu }
$ develops some constant background value

\begin{equation}
<A_{\mu }(x)>\text{ }=n_{\mu }M  \label{vev1}
\end{equation}%
and the Lorentz symmetry $SO(1,3)$ formally breaks down to $SO(3)$ or $%
SO(1,2)$ depending on the time-like ($n^{2}>0$) or space-like ($n^{2}<0$)
nature of SLIV. The point is, however, that, in sharp contrast to the
nonlinear $\sigma $ model for pions, the nonlinear QED theory, due to the
starting gauge invariance involved, ensures that all the physical Lorentz
violating effects turn out to be non-observable. It was shown \cite{nambu},
while only in the tree approximation and for the time-like SLIV ($n^{2}>0$),
that the nonlinear constraint (\ref{const}) implemented into the standard
QED Lagrangian containing a charged\ fermion field $\psi (x)$

\begin{equation}
L_{QED}=-\frac{1}{4}F_{\mu \nu }F^{\mu \nu }+\overline{\psi }(i\gamma
\partial +m)\psi -eA_{\mu }\overline{\psi }\gamma ^{\mu }\psi \text{ \ \ }
\label{lag11}
\end{equation}%
as a supplementary condition appears in fact as a possible gauge choice for
the vector field $A_{\mu }$, while the $S$-matrix remains unaltered under
such a gauge convention. Really, this nonlinear QED contains a plethora of
Lorentz and $CPT$ violating couplings when it is expressed in terms of the
pure Goldstonic photon modes ($a_{\mu }$) according to the constraint
condition (\ref{const})

\begin{equation}
A_{\mu }=a_{\mu }+\frac{n_{\mu }}{n^{2}}(M^{2}-n^{2}a^{2})^{\frac{1}{2}}%
\text{ , \ }n_{\mu }a_{\mu }=0\text{ \ \ \ \ (}a^{2}\equiv a_{\mu }a^{\mu }%
\text{).}  \label{gol}
\end{equation}%
(for definiteness, one takes the positive sign for the square root when
expanding it in powers of $a^{2}/M^{2}$). However, the contributions of all
these Lorentz violating couplings to physical processes completely cancel
out among themselves. So, SLIV is shown to be superficial as it affects only
the gauge of the vector potential $A_{\mu }$ at least in the tree
approximation \cite{nambu}.

Some time ago, this result was extended to the one-loop approximation and
for both\ the time-like ($n^{2}>0$) and space-like ($n^{2}<0$) Lorentz
violation \cite{az}. All the contributions to the photon-photon,
photon-fermion and fermion-fermion interactions violating physical Lorentz
invariance happen to exactly cancel among themselves in the manner observed
long ago by Nambu for the simplest tree-order diagrams. This means that the
constraint (\ref{const}), having been treated as a nonlinear gauge choice at
the tree (classical) level, remains as a gauge condition when quantum
effects are taken into account as well. So, in accordance with Nambu's
original conjecture, one can conclude that physical Lorentz invariance is
left intact at least in the one-loop approximation, provided we consider the
standard gauge invariant QED Lagrangian (\ref{lag11}) taken in flat
Minkowski space-time. Later this result was also confirmed for spontaneously
broken massive QED \cite{kep} (some interesting aspects of the SLIV
conditioned nonlinear QED were also considered in \cite{ur}).  It was
further argued  \cite{jej} that non-Abelian gauge fields can  also be
treated as the pseudo-Goldstone vector bosons caused by SLIV which
presumably evolves in a general Yang-Mills type theory with the nonlinear
vector field constraint $Tr(\boldsymbol{A}_{\mu }\boldsymbol{A}^{\mu })=\pm
M^{2}$\ ($M$ is a proposed SLIV scale) put on the vector field multiplet. 
Specifically, it was shown that in a theory with an internal symmetry group $%
G$ having $D$ generators not only the pure Lorentz symmetry $SO(1,3)$, but
the larger accidental symmetry $SO(D,3D)$ of the SLIV constraint in itself
appears to be spontaneously broken as well. As a result, although the pure
Lorentz violation on its own still generates only one genuine Goldstone
vector boson, the accompanying pseudo-Goldstone vector bosons related to the 
$SO(D,3D)$ breaking also come into play properly completing the whole gauge
multiplet of the internal symmetry group $G$ taken. Remarkably, they appear
to be strictly massless as well, being protected by the starting non-Abelian
gauge invariance of the Yang-Mills theory involved. When expressed in terms
of the pure Goldstone vector modes, this theory look essentially nonlinear
and contains a variety of Lorentz and $CPT$ violating couplings. However,
they do not lead to physical SLIV effects which turn out to be strictly
cancelled in all the lowest order processes considered. Actually, these
Goldstonic non-Abelian theories are in fact theories which provide the
building blocks for the Standard Model and beyond, whether they be exact as
in quantum chromodynamics or spontaneously broken as in grand unified
theories and family symmetry models \cite{ram,su3}.

Continuing this study, we here use a similar nonlinear constraint for a
symmetric two-index tensor field

\begin{equation}
H_{\mu \nu }^{2}=\mathfrak{n}^{2}M^{2}\text{ , \ \ \ \ }\mathfrak{n}%
^{2}\equiv \mathfrak{n}_{\mu \nu }\mathfrak{n}^{\mu \nu }=\pm 1\text{ }
\label{const3}
\end{equation}%
(where $\mathfrak{n}_{\mu \nu }$ is now a properly oriented `unit' Lorentz
tensor, while $M$ is the proposed scale for Lorentz violation) which fixes
its length in a similar way to the vector field case above. Also, in analogy
to the nonlinear QED case \cite{nambu} with its gauge invariant Lagrangian (%
\ref{lag11}), we propose the linearized Einstein-Hilbert kinetic term for
the tensor field, which by itself preserves a diff invariance. We show that
such a SLIV pattern (\ref{const3}), due to which the true vacuum in the
theory is chosen, induces massless tensor Goldstone modes some of which can
naturally be collected in the physical graviton. The linearized theory we
start with becomes essentially nonlinear, when expressed in terms of the
pure Goldstone modes, and contains a variety of \ Lorentz (and $CPT$)
violating couplings. However, all SLIV effects turn out to be strictly
cancelled in physical processes once the tensor field gravity theory (being
considered as the weak-field limit of general relativity (GR)) is properly
extended to GR. So, this formulation of SLIV seems to amount to the fixing
of a gauge for the tensor field in a special manner making the Lorentz
violation only superficial just as in the nonlinear QED framework \cite%
{nambu}. From this viewpoint, both conventional QED and GR theories appear
to be generic Goldstonic theories in which some of the gauge degrees of
freedom of these fields are condensed (thus eventually emerging as a
non-covariant gauge choice), while their massless NG modes are collected in
photons or gravitons in such a way that the physical Lorentz invariance is
ultimately preserved. \ However, there might appear some principal
distinctions between conventional and\ Goldstonic theories if, as we argue
later, the underlying local symmetry were slightly broken at very small
distances controlled by quantum gravity in an explicit, rather than
spontaneous, way that could eventually allow one to differentiate between
them observationally.  

The paper is organized in the following way. In section 2 we formulate the
model for tensor field gravity and find massless NG modes some of which are
collected in the physical graviton. Then in section 3 we derive general
Feynman rules for the basic graviton-graviton and graviton-matter (scalar)
field interactions in the Goldstonic gravity theory. In essence the model
contains two perturbative parameters, the inverse Planck and SLIV mass
scales, $1/M_{P}$ and $1/M,$ respectively, so that the SLIV interactions are
\ always proportional to some powers of them. Some lowest order SLIV
processes, such as graviton-graviton scattering and graviton scattering off
the massive scalar field, are considered in detail. We show that all these
Lorentz violating effects, taken in the tree approximation, in fact turn out
to vanish so that physical Lorentz invariance is ultimately restored.
Finally, in section 4 we present a resume and conclude.

\section{The Model}

According to our philosophy, we propose to consider the tensor field gravity
theory which mimics linearized general relativity in Minkowski space-time.
The corresponding Lagrangian for one real scalar field $\phi $ (representing
all sorts of matter in the model)

\begin{equation}
\mathcal{L}(H_{\mu \nu },\phi )=\mathcal{L}(H)+\mathcal{L}(\phi )+\mathcal{L}%
_{int}  \label{tl}
\end{equation}%
consists of the tensor field kinetic\ terms of the form

\begin{equation}
\mathcal{L}(H)=\frac{1}{2}\partial _{\lambda }H^{\mu \nu }\partial ^{\lambda
}H_{\mu \nu }-\frac{1}{2}\partial _{\lambda }H_{tr}\partial ^{\lambda
}H_{tr}-\partial _{\lambda }H^{\lambda \nu }\partial ^{\mu }H_{\mu \nu
}+\partial ^{\nu }H_{tr}\partial ^{\mu }H_{\mu \nu }\text{ ,}  \label{fp}
\end{equation}%
($H_{tr}$ stands for the trace of $H_{\mu \nu },$ $H_{tr}=\eta ^{\mu \nu
}H_{\mu \nu }$) which is invariant under the diff transformations

\begin{equation}
\delta H_{\mu \nu }=\partial _{\mu }\xi _{\nu }+\partial _{\nu }\xi _{\mu }%
\text{ , \ \ \ }\delta x^{\mu }=-\xi ^{\mu }(x)\text{ ,}  \label{tr3}
\end{equation}%
together with the free scalar field and interaction terms

\begin{equation}
\mathcal{L}(\phi )=\frac{1}{2}\left( \partial _{\rho }\phi \partial ^{\rho
}\phi -m^{2}\phi ^{2}\right) \text{ },\text{ \ \ \ \ }\mathcal{L}_{int}=-%
\frac{1}{M_{P}}H_{\mu \nu }T^{\mu \nu }(\phi )\text{ . \ \ \ \ \ }
\label{fh}
\end{equation}%
Here $T^{\mu \nu }(\phi )$ is the conventional energy-momentum tensor for a
scalar field

\begin{equation}
T^{\mu \nu }(\phi )=\partial ^{\mu }\phi \partial ^{\nu }\phi -\eta ^{\mu
\nu }\mathcal{L}(\phi )\text{ ,}  \label{tt}
\end{equation}%
and the coupling constant in $\mathcal{L}_{int}$ is chosen to be the inverse
of the Planck mass $M_{P}$. It is clear that, in contrast to the tensor
field kinetic \ terms, the other terms in (\ref{tl}) are only approximately
invariant under the diff transformations (\ref{tr3}) and 
\begin{equation}
\delta \phi =\xi ^{\rho }\partial _{\rho }\phi   \label{phi1}
\end{equation}%
for tensor and scalar fields, as they correspond to the weak-field limit in
GR. Following the nonlinear $\sigma $-model for QED \cite{nambu}, we propose
the SLIV condition (\ref{const3}) as some tensor field length-fixing
constraint which is supposed to be substituted into the total Lagrangian $%
\mathcal{L}(H_{\mu \nu },\phi )$ prior to the variation of the action. This
eliminates, as we will see, a massive Higgs mode in the final theory thus
leaving only massless Goldstone modes, some of which are then collected in
the physical graviton.

Let us first turn to the spontaneous Lorentz violation itself, which is
caused by the constraint (\ref{const3}). This constraint can be written in
the more explicit form

\begin{equation}
H_{\mu \nu }^{2}=H_{00}^{2}+H_{i=j}^{2}+(\sqrt{2}H_{i\neq j})^{2}-(\sqrt{2}%
H_{0i})^{2}=\mathfrak{n}^{2}M^{2}=\pm \text{ }M^{2}\text{ \ \ \ \ \ \ \ }
\label{c4}
\end{equation}%
(where the summation over all indices $(i,j=1,2,3)$ is imposed) and means in
essence that the tensor field $H_{\mu \nu }$ develops the vacuum expectation
value (vev) configuration

\begin{equation}
<H_{\mu \nu }(x)>\text{ }=\mathfrak{n}_{\mu \nu }M  \label{v}
\end{equation}%
determined by the matrix $\mathfrak{n}_{\mu \nu }$. The initial Lorentz
symmetry $SO(1,3)$ of the Lagrangian $\mathcal{L}(H_{\mu \nu },\phi )$ given
in (\ref{tl}) then formally breaks down at a scale $M$ to one of its
subgroups. If one assumes a "minimal" vacuum configuration in the $SO(1,3)$
space with the vev (\ref{v}) developed on a single $H_{\mu \nu }$ component,
there are in fact the following three possibilities

\begin{eqnarray}
(a)\text{ \ \ \ \ }\mathfrak{n}_{00} &\neq &0\text{ , \ \ }%
SO(1,3)\rightarrow SO(3)  \notag \\
(b)\text{ \ \ \ }\mathfrak{n}_{i=j} &\neq &0\text{ , \ \ }SO(1,3)\rightarrow
SO(1,2)  \label{ns} \\
(c)\text{ \ \ \ }\mathfrak{n}_{i\neq j} &\neq &0\text{ , \ \ }%
SO(1,3)\rightarrow SO(1,1)  \notag
\end{eqnarray}%
for the positive sign in (\ref{c4}), and

\begin{equation}
(d)\text{ \ \ \ }\mathfrak{n}_{0i}\neq 0\text{ , \ \ }SO(1,3)\rightarrow
SO(2)  \label{nss}
\end{equation}%
for the negative sign. These breaking channels can be readily derived by
counting how many different eigenvalues the vev matrix $\mathfrak{n}$ has
for each particular case ($a$-$d$). Accordingly, there are only three
Goldstone modes in the cases ($a,b$) and five modes in the cases ($c$-$d$) 
\footnote{%
Indeed, the vev matrices in the cases ($a,b$) look, respectively, as $%
\mathfrak{n}^{(a)}=diag(1,0,0,0)$ and $\mathfrak{n}^{(b)}=diag(0,1,0,0)$,
while in the cases ($c$-$d$) these matrices, taken in the diagonal bases,
have the forms $\mathfrak{n}^{(c)}=diag(0,1,-1,0)$ \ and $\mathfrak{n}%
^{(d)}=diag(1,-1,0,0)$, respectively (for certainty, we fixed $i=j=1$ in the
case ($b$), $i=1$ and $j=2$ in the case ($c$), and $i=1$ in the case ($d$)).
The groups of invariance of these vev matrices are just the surviving
Lorentz subgroups indicated on the right-handed sides in (\ref{ns}) and (\ref%
{nss}). The broken Lorentz generators determine then the numbers of
Goldstone modes mentioned above.}. In order to associate at least one of the
two transverse polarization states of the physical graviton with these
modes, one could have any of the above-mentioned SLIV channels except for
the case ($a$). Indeed, it is impossible for the graviton to have all
vanishing spatial components, as one needs for the Goldstone modes in case ($%
a$). Therefore, no linear combination of the three Goldstone modes in case ($%
a$) could behave like the physical graviton (see more detailed consideration
in \cite{car}). Apart from the minimal vev configuration, there are many
others as well. A particular case of interest is that of the traceless vev
tensor $\mathfrak{n}_{\mu \nu }$

\begin{equation}
\text{\ \ }\mathfrak{n}_{\mu \nu }\eta ^{\mu \nu }=0  \label{tll}
\end{equation}%
in terms of which the Goldstonic gravity Lagrangian acquires an especially
simple form (see below). It is clear that the vev in this case can be
developed on several $H_{\mu \nu }$ components simultaneously, which in
general may lead to total Lorentz violation with all six Goldstone modes
generated. For simplicity, we will use this form of vacuum configuration in
what follows, while our arguments can be applied to any type of vev tensor $%
\mathfrak{n}_{\mu \nu }$.

Aside from the pure Lorentz Goldstone modes, the question of the other
components of the symmetric two-index tensor $H_{\mu \nu }$ naturally
arises. Remarkably, they turn out to be Pseudo Goldstone modes (PGMs) in the
theory, just as it appears in the SLIV condioned Yang-Mills theory \cite{jej}%
. Indeed, although we only propose Lorentz invariance of the Lagrangian $%
\mathcal{L}(H_{\mu \nu },\phi )$, the SLIV constraint (\ref{const3})
formally possesses the much higher accidental symmetry $SO(7,3)$ of the
constrained bilinear form (\ref{c4}), which manifests itself when
considering the $H_{\mu \nu }$ components as the "vector" ones under $SO(7,3)
$. This symmetry is in fact spontaneously broken, side by side with Lorentz
symmetry, at the scale $M.$ Assuming again a minimal vacuum configuration in
the $SO(7,3)$ \ space, with the vev (\ref{v}) developed on a single $H_{\mu
\nu }$ component, we have either time-like ($SO(7,3)$ $\rightarrow SO(6,3)$)
or space-like ($SO(7,3)$ $\rightarrow SO(7,2)$) violations of the accidental
symmetry depending on the sign of $\mathfrak{n}^{2}=\pm 1$ in (\ref{c4}).
According to the number of broken $SO(7,3)$ generators, just nine massless
NG modes appear in both cases. Together with an effective Higgs component,
on which the vev is developed, they complete the whole ten-component
symmetric tensor field $H_{\mu \nu }$ of the basic Lorentz group. Some of
them are true Goldstone modes of the spontaneous Lorentz violation, others
are PGMs since, as was mentioned, an accidental $SO(7,3)$ symmetry is not
shared by the whole Lagrangian $\mathcal{L}(H_{\mu \nu },\phi )$ given in (%
\ref{tl}). Notably, in contrast to the scalar PGM case \cite{GL}, they
remain strictly massless being protected by the starting diff invariance%
\footnote{%
For non-minimal vacuum configuration when vevs are developed on several \ $%
H_{\mu \nu }$\ components, thus leading to a more substantial breaking of
the accidental $SO(7,3)$ symmetry, some extra PGMs are also generated.
However, they are not protected by a diff invariance and acquire masses of
the order of the breaking scale $M$.} which becomes exact when the tensor
field gravity Lagrangian (\ref{tl}) is properly extended to GR. Owing to
this invariance, some of the Lorentz Goldstone modes and PGMs can then be
gauged away from the theory, as usual.

Now, one can rewrite the Lagrangian $\mathcal{L}(H_{\mu \nu },\phi )$ in
terms of the Goldstone modes explicitly using the SLIV constraint (\ref%
{const3}). For this purpose, let us take the following handy
parameterization for the tensor field $H_{\mu \nu }$

\begin{equation}
H_{\mu \nu }=h_{\mu \nu }+\frac{n_{\mu \nu }}{n^{2}}(\mathfrak{n}\cdot
H)\qquad (\mathfrak{n}\cdot H\equiv \mathfrak{n}_{\mu \nu }H^{\mu \nu })
\label{par}
\end{equation}%
where $h_{\mu \nu }$ corresponds to the pure Goldstonic modes \footnote{%
It should be particularly emphasized that the modes collected in the $h_{\mu
\nu }$ are in fact the Goldstone modes of the broken accidental $SO(7,3)$
symmetry of the constraint (\ref{const3}), thus containing the Lorentz
Goldstone modes and PGMs put together.} satisfying

\begin{equation}
\text{\ }\mathfrak{n}\cdot h=0\text{\ }\qquad (\mathfrak{n}\cdot h\equiv 
\mathfrak{n}_{\mu \nu }h^{\mu \nu })  \label{sup}
\end{equation}%
while the effective \textquotedblleft Higgs" mode (or the $H_{\mu \nu }$
component in the vacuum direction) is given by the scalar product $\mathfrak{%
n}\cdot H$. Substituting this parameterization (\ref{par}) into the tensor
field constraint (\ref{const3}), one comes to the equation for $\mathfrak{n}%
\cdot H$

\begin{equation}
\text{\ }\mathfrak{n}\cdot H\text{\ }=(M^{2}-\mathfrak{n}^{2}h^{2})^{\frac{1%
}{2}}=M-\frac{\mathfrak{n}^{2}h^{2}}{2M}+O(1/M^{2})  \label{constr1}
\end{equation}%
taking, for definiteness, the positive sign for the square root and
expanding it in powers of $h^{2}/M^{2}$, $h^{2}\equiv h_{\mu \nu }h^{\mu \nu
}$. Putting then the parameterization (\ref{par}) with the SLIV constraint (%
\ref{constr1}) into the Lagrangian $\mathcal{L}(H_{\mu \nu },\phi )$ given
in (\ref{tl}, \ref{fp}, \ref{fh}), one comes to the truly Goldstonic tensor
field gravity Lagrangian $\mathcal{L}(h_{\mu \nu },\phi )$ containing an
infinite series in powers of the $h_{\mu \nu }$ modes. For the traceless vev
tensor $\mathfrak{n}_{\mu \nu }$ (\ref{tll}) it takes, without loss of
generality, the especially simple form \ \ \ \ 

\begin{eqnarray}
\mathcal{L}(h_{\mu \nu },\phi ) &=&\frac{1}{2}\partial _{\lambda }h^{\mu \nu
}\partial ^{\lambda }h_{\mu \nu }-\frac{1}{2}\partial _{\lambda
}h_{tr}\partial ^{\lambda }h_{tr}-\partial _{\lambda }h^{\lambda \nu
}\partial ^{\mu }h_{\mu \nu }+\partial ^{\nu }h_{tr}\partial ^{\mu }h_{\mu
\nu }+  \label{gl} \\
&&+\frac{1}{2M}h^{2}\left[ -2\mathfrak{n}^{\mu \lambda }\partial _{\lambda
}\partial ^{\nu }h_{\mu \nu }+\mathfrak{n}^{2}(\mathfrak{n}\partial \partial
)h_{tr}\right] +\frac{1}{8M^{2}}h^{2}\left[ -\mathfrak{n}^{2}\partial
^{2}+2(\partial \mathfrak{nn}\partial )\right] h^{2}  \notag \\
&&+\mathcal{L}(\phi )-\frac{M}{M_{P}}\mathfrak{n}^{2}\left[ \mathfrak{n}%
_{\mu \nu }\partial ^{\mu }\phi \partial ^{\nu }\phi \right] -\frac{1}{M_{P}}%
h_{\mu \nu }T^{\mu \nu }-\frac{1}{2MM_{P}}h^{2}\left[ -\mathfrak{n}_{\mu \nu
}\partial ^{\mu }\phi \partial ^{\nu }\phi \right]   \notag
\end{eqnarray}%
written in the $O(h^{2}/M^{2})$ approximation in which, besides the
conventional graviton bilinear kinetic terms, there are also three- and
four-linear interaction terms in powers of $h_{\mu \nu }$ in the Lagrangian.
Some of the notations used are collected below

\begin{eqnarray}
h^{2} &\equiv &h_{\mu \nu }h^{\mu \nu }\text{ , \ \ }h_{tr}\equiv \eta ^{\mu
\nu }h_{\mu \nu }\text{ , \ }  \label{n} \\
\mathfrak{n}\partial \partial &\equiv &\mathfrak{n}_{\mu \nu }\partial ^{\mu
}\partial ^{\nu }\text{\ , \ \ }\partial \mathfrak{nn}\partial \equiv
\partial ^{\mu }\mathfrak{n}_{\mu \nu }\mathfrak{n}^{\nu \lambda }\partial
_{\lambda }\text{ .\ \ \ }  \notag
\end{eqnarray}

The bilinear scalar field term%
\begin{equation}
-\frac{M}{2M_{P}}\mathfrak{n}^{2}\left[ \mathfrak{n}_{\mu \nu }\partial
^{\mu }\phi \partial ^{\nu }\phi \right]   \label{ttt}
\end{equation}%
in the third line in the Lagrangian (\ref{gl}) merits special notice. This
term arises from the interaction Lagrangian $\mathfrak{L}_{int}$ (\ref{fh})
after application of the tracelessness condition (\ref{tll}) for the vev
tensor $\mathfrak{n}_{\mu \nu }$. It could significantly affect the
dispersion relation for the scalar field $\phi $ (and any other sort of
matter as well), thus leading to an unacceptably large Lorentz violation if
the SLIV scale $M$ were comparable with the Planck mass $M_{P}.$ However,
this term can be gauged away by an appropriate choice of the gauge parameter
function $\xi ^{\mu }(x)$ in the transformations (\ref{tr3}, \ref{phi1}) of
the tensor and scalar fields\footnote{%
Actually, in the Lagrangian $\mathcal{L}(H_{\mu \nu },\phi )$ the vacuum
shift of the tensor field $H_{\mu \nu }=h_{\mu \nu }+\frac{\mathfrak{n}_{\mu
\nu }}{\mathfrak{n}^{2}}M$ is in fact equivalent to a gauge transformation
which, for the appropriately chosen transformation of the scalar field $\phi
(x),$ leaves the corresponding action invariant.}. Technically, one simply
transforms the scalar field and its derivative to a new coordinate system $%
x^{\mu }\rightarrow $ $x^{\mu }-\xi ^{\mu }$ in the Goldstonic Lagrangian $%
\mathcal{L}(h_{\mu \nu },\phi )$. Actually, using the fixed-point variation
of $\phi (x)$ given above in (\ref{phi1}) and differentiating both sides
with respect to $x^{\mu }$ one obtains 
\begin{equation}
\delta (\partial _{\mu }\phi )=\partial _{\mu }(\xi ^{\nu }\partial _{\nu
}\phi ).
\end{equation}%
This gives in turn 
\begin{equation}
\delta _{tot}(\partial _{\mu }\phi )=\delta (\partial _{\mu }\phi )+\delta
x^{\nu }\partial _{\nu }(\partial _{\mu }\phi )=\partial _{\mu }\xi ^{\nu
}\partial _{\nu }\phi   \label{redd}
\end{equation}%
for the total variation of the scalar field derivative. The corresponding
total variation of the Goldstonic tensor $h_{\mu \nu }$, caused by the same
transformation to the coordinate system $x^{\mu }-\xi ^{\mu }$, is given in
turn by equations (\ref{tr3}) and (\ref{par}) to be 
\begin{equation}
\delta _{tot}h_{\mu \nu }=(\partial ^{\rho }\xi ^{\sigma }+\partial ^{\sigma
}\xi ^{\rho })\left( \eta _{\rho \mu }\eta _{\sigma \nu }-\frac{\mathfrak{n}%
_{\mu \nu }}{\mathfrak{n}^{2}}\mathfrak{n}_{\rho \sigma }\right) -\xi ^{\rho
}\partial _{\rho }h_{\mu \nu }.  \label{dh}
\end{equation}%
One can now readily see that, with the parameter function $\xi ^{\mu }(x)$
chosen as 
\begin{equation}
\xi ^{\mu }(x)=\frac{M}{2M_{P}}\mathfrak{n}^{2}\mathfrak{n}^{\mu \nu }x_{\nu
}\text{ },
\end{equation}%
the dangerous term (\ref{ttt}) is precisely cancelled\footnote{%
In the general case, with the vev tensor $\mathfrak{n}_{\mu \nu }$ having a
non-zero trace, this cancellation would also require the redefinition of the
scalar field itself as $\phi \rightarrow \phi (1-\mathfrak{n}_{\mu \nu }\eta
^{\mu \nu }\frac{M}{M_{P}})^{-1/2}$.} by an analogous term stemming from the
scalar field kinetic term in the $\mathfrak{L}(\phi )$ given in (\ref{fh}),
while the total variation of the tensor $h_{\mu \nu }$ reduces to just the
second term in (\ref{dh}). This term is of the natural order $O(\xi h)$,
which can be neglected in the weak field approximation, so that to the
present accuracy the tensor field variation $\delta _{tot}h_{\mu \nu }=0$.
Indeed, since the diff invariance is an approximate symmetry of the
Lagrangian $\mathcal{L}(h_{\mu \nu },\phi )$, the above cancellation will
only be accurate up to the order corresponding to the linearized Lagrangian $%
\mathcal{L}(H_{\mu \nu },\phi )$ we started with in (\ref{tl}). Actually, a
proper extension of the tensor field theory to GR with its exact diff
invariance will ultimately restore the usual form of the dispersion relation
for the scalar (and other matter) fields. Taking this into account, we will
henceforth omit the term (\ref{ttt}) in $\mathcal{L}(h_{\mu \nu },\phi )$,
thus keeping the \textquotedblleft normal" dispersion relation for the
scalar field in what follows.

Together with the Lagrangian one must also specify other supplementary
conditions for the tensor field $h^{\mu \nu }$(appearing eventually as
possible gauge fixing terms in the Goldstonic tensor field gravity) in
addition to the basic Goldstonic \ "gauge" \ condition $\mathfrak{n}_{\mu
\nu }h^{\mu \nu }=0$ given above (\ref{sup}). The point is that the spin $1$
states are still left in the theory and are described by some of the
components of \ the new tensor $h_{\mu \nu }.$ This is certainly
inadmissible \footnote{%
Indeed, spin $1$ must be necessarily excluded as the sign of the energy for
spin $1$ is always opposite to that for spin $2$ and $0.$}. Usually, the
spin $1$ states (and one of the spin $0$ states) are excluded by the
conventional Hilbert-Lorentz condition

\begin{equation}
\partial ^{\mu }h_{\mu \nu }+q\partial _{\nu }h_{tr}=0  \label{HL}
\end{equation}%
($q$ is an arbitrary constant, giving for $q=-1/2$ the standard harmonic
gauge condition). However, as we have already imposed the constraint (\ref%
{sup}), we can not use the full Hilbert-Lorentz condition (\ref{HL})
eliminating four more degrees of freedom in $h_{\mu \nu }.$ Otherwise, we
would have an "over-gauged" theory with a non-propagating graviton. In fact,
the simplest set of conditions which conform with the Goldstonic condition (%
\ref{sup}) turns out to be

\begin{equation}
\partial ^{\rho }(\partial _{\mu }h_{\nu \rho }-\partial _{\nu }h_{\mu \rho
})=0  \label{gauge}
\end{equation}%
This set excludes only three degrees of freedom \footnote{%
The solution for a gauge function $\xi _{\mu }(x)$ satisfying the condition$%
\ $(\ref{gauge}) $\ $\ can generally be chosen as $\xi _{\mu }=$\ $\ \square
^{-1}(\partial ^{\rho }h_{\mu \rho })+\partial _{\mu }\theta $, where $%
\theta (x)$ is an arbitrary scalar function, so that only three degrees of
freedom in $h_{\mu \nu }$ are actually eliminated.} in $h_{\mu \nu }$ and,
besides, it automatically satisfies the Hilbert-Lorentz spin condition as
well. So, with the Lagrangian (\ref{gl}) and the supplementary conditions\ (%
\ref{sup}) and (\ref{gauge}) lumped together, one eventually comes to a
working model for the Goldstonic tensor field gravity. Generally, from ten
components of the symmetric two-index tensor $h_{\mu \nu }$ four components
are excluded by the supplementary conditions (\ref{sup}) and (\ref{gauge}).
For a plane gravitational wave propagating in, say, the $z$ direction
another four components are also eliminated, due to the fact that the above
supplementary conditions still leave freedom in the choice of a coordinate
system, $x^{\mu }\rightarrow $ $x^{\mu }-\xi ^{\mu }(t-z/c),$ much as it
takes place in standard GR. Depending on the form of the vev tensor $%
\mathfrak{n}_{\mu \nu }$, caused by SLIV, the two remaining transverse modes
of the physical graviton may consist solely of Lorentz Goldstone modes or of
Pseudo Goldstone modes, or include both of them.

\section{The Lowest Order SLIV Processes}

The Goldstonic gravity Lagrangian (\ref{gl}) looks essentially nonlinear and
contains a variety of \ Lorentz and $CPT$ violating couplings when expressed
in terms of the pure tensor Goldstone modes. However, as we show below,\ all
violation effects turn out to be strictly cancelled in the lowest order SLIV
processes. Such a cancellation in vector-field theories, both Abelian \cite%
{nambu,az,kep} and non-Abelian \cite{jej}, and, therefore, their equivalence
to conventional QED and Yang-Mills theories, allows one to conclude that the
nonlinear SLIV constraint in these theories amounts to a non-covariant gauge
choice in an otherwise gauge invariant and Lorentz invariant theory. It
seems that a similar conclusion can be made for tensor field gravity, i.e.
the SLIV constraint (\ref{const3}) corresponds to a special gauge choice in
a diff and Lorentz invariant theory. This conclusion certainly works for the
diff invariant free tensor field part (\ref{fp}) in the starting Lagrangian $%
\mathcal{L}(H_{\mu \nu },\phi )$. On the other hand, its matter field sector
(\ref{fh}), possessing only an approximate diff invariance, might lead to an
actual Lorentz violation through the deformed dispersion relations of the
matter fields involved. However, as was mentioned above, a proper extension
of the tensor field theory to GR with its exact diff invariance ultimately
restores the dispersion relations for matter fields and, therefore, the SLIV
effects vanish. Taking this into account, we omit the term (\ref{ttt}) in
the\ Goldstonic gravity Lagrangian $\mathcal{L}(h_{\mu \nu },\phi )$ thus
keeping the "normal" dispersion relation for the scalar field representing
all the matter in our model.

We are now going to consider the lowest order SLIV processes, after first
establishing the Feynman rules in the Goldstonic gravity theory. We use for
simplicity, both in the Lagrangian $\mathcal{L}$ (\ref{gl}) and forthcoming
calculations, the traceless vev tensor $\mathfrak{n}_{\mu \nu }$, while our
results remain true for any type of vacuum configuration caused by SLIV.

\subsection{Feynman rules}

The Feynman rules stemming from the Lagrangian $\mathcal{L}$ (\ref{gl}) for
the pure graviton sector are as follows:

(\textbf{i}) The first and most important is the graviton propagator which
only conforms with the Lagrangian (\ref{gl}) and the gauge conditions\ (\ref%
{sup}) and (\ref{gauge})

\begin{eqnarray}
-iD_{\mu \nu \alpha \beta }\left( k\right) &=&\frac{1}{2k^{2}}\left( \eta
_{\beta \mu }\eta _{\alpha \nu }+\eta _{\beta \nu }\eta _{\alpha \mu }-\eta
_{\alpha \beta }\eta _{\mu \nu }\right)  \notag \\
&&-\frac{1}{2k^{4}}\left( \eta _{\beta \nu }k_{\alpha }k_{\mu }+\eta
_{\alpha \nu }k_{\beta }k_{\mu }+\eta _{\beta \mu }k_{\alpha }k_{\nu }+\eta
_{\alpha \mu }k_{\beta }k_{\nu }\right)  \label{prop} \\
&&-\frac{1}{k^{2}(\mathfrak{n}kk)}\left( k_{\alpha }k_{\beta }\mathfrak{n}%
_{\mu \nu }+k_{\nu }k_{\mu }\mathfrak{n}_{\alpha \beta }\right) +\frac{1}{%
k^{2}(\mathfrak{n}kk)^{2}}\left[ \mathfrak{n}^{2}-\frac{2}{k^{2}}(k\mathfrak{%
nn}k)\right] k_{\mu }k_{\nu }k_{\alpha }k_{\beta }  \notag \\
&&+\frac{1}{k^{4}(\mathfrak{n}kk)}\left( \mathfrak{n}_{\mu \rho }k^{\rho
}k_{\nu }k_{\alpha }k_{\beta }+\mathfrak{n}_{\nu \rho }k^{\rho }k_{\mu
}k_{\alpha }k_{\beta }+\mathfrak{n}_{\alpha \rho }k^{\rho }k_{\nu }k_{\mu
}k_{\beta }+\mathfrak{n}_{\beta \rho }k^{\rho }k_{\nu }k_{\alpha }k_{\mu
}\right)  \notag
\end{eqnarray}%
(where $(\mathfrak{n}kk)\equiv \mathfrak{n}_{\mu \nu }k^{\mu }k^{\nu }$ and $%
(k\mathfrak{nn}k)\equiv k^{\mu }\mathfrak{n}_{\mu \nu }\mathfrak{n}^{\nu
\lambda }k_{\lambda })$. It automatically satisfies the orthogonality
condition $\mathfrak{n}^{\mu \nu }D_{\mu \nu \alpha \beta }\left( k\right)
=0 $ and on-shell transversality $k^{\mu }k^{\nu }D_{\mu \nu \alpha \beta
}(k,k^{2}=0)=0.$ $\ $This is consistent with the corresponding polarization
tensor $\epsilon _{\mu \nu }(k,k^{2}=0)$ of the free tensor fields, being
symmetric, traceless ($\eta ^{\mu \nu }\epsilon _{\mu \nu }=0),$ transverse (%
$k^{\mu }\epsilon _{\mu \nu }=0),$ and also orthogonal to the vacuum
direction, $\mathfrak{n}^{\mu \nu }\epsilon _{\mu \nu }(k)=0$. Apart from
that, the gauge invariance allows us to write the polarization tensor in the
factorized form \cite{gross} , $\epsilon _{\mu \nu }(k)=\epsilon _{\mu
}(k)\epsilon _{\nu }(k),$ and to proceed with the above-mentioned
tracelessness and transversality expressed as the simple conditions $%
\epsilon _{\mu }\epsilon ^{\mu }=0$ and $k^{\mu }\epsilon _{\mu }=0$
respectively. In the following we will use these simplifications. As one can
see, only the standard terms given by the first bracket in (\ref{prop})
contribute when the propagator is sandwiched between conserved
energy-momentum tensors of matter fields, and the result is always Lorentz
invariant.

(\textbf{ii}) Next is the 3-graviton vertex $\ $with graviton polarization
tensors (and 4-momenta) given by $\epsilon ^{\alpha \alpha ^{\prime
}}(k_{1}),$ $\epsilon ^{\beta \beta ^{\prime }}(k_{2})$ and $\epsilon
^{\gamma \gamma ^{\prime }}(k_{3})$

\begin{eqnarray}
&&-\frac{i}{2M}P^{\alpha \alpha ^{\prime }}(k_{1})\left( \eta ^{\beta \gamma
}\eta ^{\beta ^{\prime }\gamma ^{\prime }}+\eta ^{\beta \gamma ^{\prime
}}\eta ^{\beta ^{\prime }\gamma }\right)  \notag \\
&&-\frac{i}{2M}P^{\beta \beta ^{\prime }}(k_{2})\left( \eta ^{\alpha \gamma
}\eta ^{\alpha ^{\prime }\gamma ^{\prime }}+\eta ^{\alpha \gamma ^{\prime
}}\eta ^{\alpha ^{\prime }\gamma }\right)  \label{3} \\
&&-\frac{i}{2M}P^{\gamma \gamma ^{\prime }}(k_{3})\left( \eta ^{\beta \alpha
}\eta ^{\beta ^{\prime }\alpha ^{\prime }}+\eta ^{\beta \alpha ^{\prime
}}\eta ^{\beta ^{\prime }\alpha }\right)  \notag
\end{eqnarray}%
where the momentum tensor $P^{\mu \nu }(k)$ is \ 

\begin{equation}
P^{\mu \nu }(k)=-\mathfrak{n}^{\nu \rho }k_{\rho }k^{\mu }-\mathfrak{n}^{\mu
\rho }k_{\rho }k^{\nu }+\eta ^{\mu \nu }\mathfrak{n}^{\rho \sigma }k_{\rho
}k_{\sigma }\text{ .}  \label{p}
\end{equation}%
Note that all 4-momenta at the vertices are taken ingoing throughout.

(\textbf{iii}) Finally, the 4-graviton vertex $\ $with the graviton
polarization tensors (and 4-momenta) $\epsilon ^{\alpha \alpha ^{\prime
}}(k_{1}),$ $\epsilon ^{\beta \beta ^{\prime }}(k_{2}),$ $\epsilon ^{\gamma
\gamma ^{\prime }}(k_{3})$ and \ $\epsilon ^{\delta \delta ^{\prime
}}(k_{4}) $

\begin{eqnarray}
&&iQ_{\mu \nu }\left( \eta ^{\alpha \beta }\eta ^{\alpha ^{\prime }\beta
^{\prime }}+\eta ^{\alpha \beta ^{\prime }}\eta ^{\alpha ^{\prime }\beta
}\right) \left( \eta ^{\gamma \delta }\eta ^{\gamma ^{\prime }\delta
^{\prime }}+\eta ^{\gamma \delta ^{\prime }}\eta ^{\gamma ^{\prime }\delta
}\right) (k_{1}+k_{2})^{\mu }(k_{1}+k_{2})^{\nu }  \notag \\
&&+iQ_{\mu \nu }\left( \eta ^{\alpha \gamma }\eta ^{\alpha ^{\prime }\gamma
^{\prime }}+\eta ^{\alpha \gamma ^{\prime }}\eta ^{\alpha ^{\prime }\gamma
}\right) \left( \eta ^{\beta \delta }\eta ^{\beta ^{\prime }\delta ^{\prime
}}+\eta ^{\beta \delta ^{\prime }}\eta ^{\beta ^{\prime }\delta }\right)
(k_{1}+k_{3})^{\mu }(k_{1}+k_{3})^{\nu }  \label{4} \\
&&+iQ_{\mu \nu }\left( \eta ^{\alpha \delta }\eta ^{\alpha ^{\prime }\delta
^{\prime }}+\eta ^{\alpha \delta ^{\prime }}\eta ^{\alpha ^{\prime }\delta
}\right) \left( \eta ^{\gamma \beta }\eta ^{\gamma ^{\prime }\beta ^{\prime
}}+\eta ^{\gamma \beta ^{\prime }}\eta ^{\gamma ^{\prime }\beta }\right)
(k_{1}+k_{4})^{\mu }(k_{1}+k_{4})^{\nu }.  \notag
\end{eqnarray}%
Here we have used the self-evident identities for all ingoing momenta ($%
k_{1}+k_{2}+k_{3}+k_{4}=0$), such as

\begin{equation*}
(k_{1}+k_{2})^{\mu }(k_{1}+k_{2})^{\nu }+(k_{3}+k_{4})^{\mu
}(k_{3}+k_{4})^{\nu }=2(k_{1}+k_{2})^{\mu }(k_{1}+k_{2})^{\nu }
\end{equation*}%
and so on, and denoted by $Q_{\mu \nu }$ the expression

\begin{equation}
Q_{\mu \nu }\equiv -\frac{1}{4M^{2}}(-\mathfrak{n}^{2}\eta _{\mu \nu }+2%
\mathfrak{n}_{\mu \rho }\mathfrak{n}_{\nu }^{\rho })\text{ .}  \label{Q}
\end{equation}%
Coming now to the gravitational interaction of the scalar field, one has two
more vertices: \ 

(\textbf{iv}) The standard graviton-scalar-scalar vertex with the graviton
polarization tensor $\epsilon ^{\alpha \alpha ^{\prime }}$ and the scalar
field 4-momenta $p_{1}$ and $p_{2}$

\begin{equation}
\frac{i}{M_{p}}\left( p_{1}^{\alpha }p_{2}^{\alpha ^{\prime }}+p_{2}^{\alpha
}p_{1}^{\alpha ^{\prime }}\right) -\frac{i}{M_{p}}\eta ^{\alpha \alpha
^{\prime }}[(p_{1}p_{2})+m^{2}]  \label{gss}
\end{equation}%
where $(p_{1}p_{2})$ stands for the scalar product.

(\textbf{v}) The contact graviton-graviton-scalar-scalar interaction caused
by SLIV \ with the graviton polarization tensors $\epsilon ^{\alpha \alpha
^{\prime }}$ and $\epsilon ^{\beta \beta ^{\prime }}$ and the scalar field
4-momenta $p_{1}$ and $p_{2}$

\begin{equation}
-\frac{i}{MM_{p}}\left( g^{\alpha \beta }g^{\alpha ^{\prime }\beta ^{\prime
}}+g^{\alpha \beta ^{\prime }}g^{\alpha ^{\prime }\beta }\right) \left( 
\mathfrak{n}_{\mu \nu }p_{1}^{\mu }p_{2}^{\nu }\right) \text{ .}
\label{ggss}
\end{equation}%
Just the rules (\textbf{i-v}) are needed to calculate the lowest order
processes mentioned above.

\subsection{Graviton-graviton scattering}

The matrix element for this SLIV process to the lowest order $1/M^{2}$ is
given by the contact $h^{4}$ vertex (\ref{4}) and the pole diagrams with
longitudinal graviton exchange between two Lorentz violating $h^{3}$
vertices (\ref{3}). There are three pole diagrams in total, describing the
elastic graviton-graviton scattering in the $s$- and $t$-channels
respectively, and also the diagram with an interchange of identical
gravitons. Remarkably, the contribution of each of them is exactly cancelled
by one of three terms appearing in the contact vertex (\ref{4}). Actually,
for the $s$-channel pole diagrams with ingoing gravitons with polarizations
(and 4-momenta) $\epsilon _{1}(k_{1})$ and $\epsilon _{2}(k_{2})$ and
outgoing gravitons with polarizations (and 4-momenta) $\epsilon _{3}(k_{3})$
and $\epsilon _{4}(k_{4})$ one has, after some \ evident simplifications
related to the graviton propagator $D_{\mu \nu }(k)$ (\ref{prop}) inside the
matrix element

\begin{equation}
i\mathcal{M}_{pole}^{(1)}=i\frac{1}{M^{2}}(\epsilon _{1}\cdot \epsilon
_{2})^{2}\left( \epsilon _{3}\cdot \epsilon _{4}\right) ^{2}(-\mathfrak{n}%
^{2}k^{2}+2k^{\mu }\mathfrak{n}_{\mu \nu }\mathfrak{n}^{\nu \lambda
}k_{\lambda }).  \label{m}
\end{equation}%
Here $k=k_{1}+k_{2}=-(k_{3}+k_{4})$ is the momentum running in the diagrams
listed above, and all the polarization tensors are properly factorized
throughout, $\epsilon _{\mu \nu }(k)=\epsilon _{\mu }(k)\epsilon _{\nu }(k),$
as was mentioned above. We have also used that, since ingoing and outgoing
gravitons appear transverse ($k_{a}^{\mu }\epsilon _{\mu }(k_{a})=0,$ $%
a=1,2,3,4$), only the third term in the momentum tensors $P^{\mu \nu
}(k_{a}) $ (\ref{p}) in the $h^{3}$ couplings (\ref{3}) contributes to all
pole diagrams. Now, one can readily confirm that this matrix element is
exactly cancelled with the first term in the contact SLIV vertex (\ref{4}),
when it is properly contracted with the graviton polarization vectors. In a
similar manner, two other terms in the contact vertex provide the further
one-to-one cancellations with the remaining two pole matrix elements $i%
\mathcal{M}_{pole}^{(2,3)}$. So, the Lorentz violating contribution to
graviton-graviton scattering is absent in Goldstonic gravity theory in the
lowest $1/M^{2}$ approximation.

\subsection{Graviton scattering on a massive scalar}

This SLIV process appears in the order $1/MM_{p}$ (in contrast to the
conventional $1/M_{p}^{2}$ order graviton-scalar scattering). It is directly
related to two diagrams one of which is given by the contact
graviton-graviton-scalar-scalar vertex (\ref{ggss}), while the other
corresponds to the pole diagram with longitudinal graviton exchange between
the Lorentz violating$\ h^{3}$ vertex (\ref{3}) and the ordinary
graviton-scalar-scalar vertex (\ref{gss}). Again, since ingoing and outgoing
gravitons appear transverse ($k_{a}^{\mu }\epsilon _{\mu }(k_{a})=0,$ $a=1,2$%
), only the third term in the momentum tensors $P^{\mu \nu }(k_{a})$ (\ref{p}%
) in the $h^{3}$ coupling (\ref{3}) contributes to this pole diagram. Apart
from that, the most crucial point is that, due to the scalar field
energy-momentum tensor conservation, the terms in the inserted graviton
propagator (\ref{prop}) other than the standard ones (first bracket in (\ref%
{prop})) give a vanishing result. Keeping all this in mind together with the
momenta satisfying $k_{1}+k_{2}+p_{1}+p_{2}=0$ ($k_{1,2}$ and $p_{1,2}$ are
the graviton and scalar field 4-momenta, respectively), one readily comes to
a simple matrix element for the pole diagram

\begin{equation}
i\mathcal{M}_{pole}=\frac{2i}{MM_{p}}\phi \left( p_{2}\right) (\varepsilon
_{1}\cdot \varepsilon _{2})^{2}\left( \mathfrak{n}_{\mu \nu }p_{1}^{\mu
}p_{2}^{\nu }\right) \phi \left( p_{1}\right) .  \label{mm}
\end{equation}%
This pole term is precisely cancelled by the contact term, $i\mathcal{M}%
_{con}$, when the SLIV vertex (\ref{ggss}) is properly contracted with the
graviton polarization vectors and the scalar boson wave functions. Again, we
may conclude that physical Lorentz invariance is left intact in graviton
scattering on a massive scalar, provided that\ its dispersion relation is
supposed to be recovered when going from the tensor field Lagrangian $%
\mathcal{L}$ (\ref{gl}) to general relativity, as was argued above.

\subsection{ Scalar-scalar scattering}

This process, due to graviton exchange, appears in the order $1/M_{P}^{2}$
and again is given by an ordinary Lorentz invariant amplitude. As was
mentioned above, only the standard terms given by the first bracket in the
graviton propagator (\ref{prop}) contribute when it is sandwiched between
conserved energy-momentum tensors of matter fields. Actually, as one can
easily confirm, the contraction of any other term in (\ref{prop}) depending
on the graviton 4-momentum $k=p_{1}+p_{2}=-(p_{3}+p_{4})$ with the
graviton-scalar-scalar vertex\ (\ref{gss})\ \ gives a zero result.

\subsection{Other processes}

Many other tree level Lorentz violating processes, related to gravitons and
scalar fields (matter fields, in general) appear in higher orders in the
basic SLIV parameter $1/M$, by iteration of couplings presented in our basic
Lagrangian (\ref{gl}) or from a further expansion of the effective Higgs
mode (\ref{constr1}) inserted into the starting Lagrangian (\ref{tl}).
Again, their amplitudes are essentially determined by an interrelation
between the longitudinal graviton exchange diagrams and the corresponding
contact multi-graviton interaction diagrams, which appear to cancel each
other, thus eliminating physical Lorentz violation in the theory.

Most likely, the same conclusion could be expected for SLIV\ loop
contributions as well. Actually, as in the massless QED case considered
earlier \cite{az}, the corresponding one-loop matrix elements in the
Goldstonic gravity theory could either vanish by themselves or amount to the
differences between pairs of similar integrals whose integration variables
are shifted relative to each other by some constants (being in general
arbitrary functions of the external four-momenta of the particles involved)
which, in the framework of dimensional regularization, could lead to their
total cancelation.

So, the Goldstonic tensor field gravity theory is likely to be physically
indistinguishable from conventional general relativity taken in the
weak-field limit, provided that the underlying diff invariance is kept
exact. This, as we have seen, requires the tensor field gravity to be
extended to GR, in order not to otherwise have an actual Lorentz violation
in the matter field sector. In this connection, the question arises whether
or not the SLIV cancellations continue to work once the tensor field gravity
theory is extended to GR, which introduces many additional terms in the
starting Lagrangian $\mathcal{L}(H_{\mu \nu },\phi )$ (\ref{tl}). \ Indeed,
since all the new terms are multi-linear in $H_{\mu \nu }$ and contain
higher orders in $1/M_{P}$, the "old" SLIV\ cancellations (considered above)
will not be disturbed, while "new" cancellations will be provided, as one
should expect, by an extended diff invariance. This extended diff invariance
follows from the proper expansion of the metric transformation law in GR

\begin{equation}
\delta g_{\mu \nu }=\partial _{\mu }\xi ^{\rho }g_{\rho \nu }+\partial _{\nu
}\xi ^{\rho }g_{\mu \rho }+\xi ^{\rho }\partial _{\rho }g_{\mu \nu }
\end{equation}%
up to the order in which the extended tensor field theory, given by the
modified Lagrangian $\mathcal{L}_{ext}(H_{\mu \nu },\phi )$, is considered.

\section{Conclusion}

We have considered spontaneous Lorentz violation, appearing through the
length-fixing tensor field constraint $H_{\mu \nu }^{2}=\pm M^{2}$ ($M$ is
the proposed scale for Lorentz violation), in the tensor field gravity
theory which mimics general relativity in Minkowski space-time. We have
shown that such a SLIV pattern, due to which the true vacuum in the theory
is chosen, induces massless tensor Goldstone modes some of which can
naturally be associated with the physical graviton. This theory looks
essentially nonlinear and contains a variety of \ Lorentz and $CPT$
violating couplings, when expressed in terms of the pure tensor Goldstone
modes. Nonetheless, all the SLIV effects turn out to be strictly cancelled
in the lowest order graviton-graviton scattering, due to the diff invariance
of the free tensor field Lagrangian (\ref{fp}) we started with. At the same
time, actual Lorentz violation may appear in the matter field interaction
sector (\ref{fh}), which only possesses an approximate diff invariance,
through deformed dispersion relations of the matter fields involved.
However, a proper extension of the tensor field theory to GR, with its exact
diff invariance, ultimately restores the normal dispersion relations for
matter fields and, therefore, the SLIV effects vanish. So, as we generally
argue, the measurable effects of SLIV, induced by elementary vector or
tensor fields, can be related to the accompanying gauge symmetry breaking
rather than to spontaneous Lorentz violation. The latter appears by itself
to be physically unobservable and only results in a non-covariant gauge
choice in an otherwise gauge invariant and Lorentz invariant theory.

From this standpoint, the only way for physical Lorentz violation to appear
would be if the above local invariance is slightly broken at very small
distances. This is in fact a place where the Goldstonic vector and tensor
field theories drastically differ from conventional QED, Yang-Mills and GR
theories. Actually, such a local symmetry breaking could lead in the former
case to deformed dispersion relations for all the matter fields involved.
This effect typically appears proportional to some power of the ratio $\frac{%
M}{M_{P}}$ (just as we have seen above for the scalar field in our model,
see (\ref{ttt})), \ though being properly suppressed by tiny gauge
non-invariance. Remarkably, the higher the SLIV scale $M$ becomes the larger
becomes the actual Lorentz violation which, for some value of the scale $M$,
may become physically observable even at low energies. Another basic
distinction of Goldstonic theories with non-exact gauge invariance is the
emergence of a mass for the graviton and other gauge fields (namely, for the
non-Abelian ones, see \cite{jej}), if they are composed from Pseudo
Goldstone modes rather than from pure Goldstone ones. Indeed, these PGMs are
no longer protected by gauge invariance and may properly acquire tiny
masses, which still do not contradict experiment. This may lead to a massive
gravity theory where the graviton mass emerges dynamically, thus avoiding
the notorious discontinuity problem \cite{zvv}. So, while Goldstonic
theories with exact local invariance are physically indistinguishable from
conventional gauge theories, there are some principal distinctions when this
local symmetry is slightly broken which could eventually allow us to
differentiate between the two types of theory in an observational way.

One could imagine how such a local symmetry breaking might occur. As was
earlier argued \cite{cj}, only local invariant theories provide the needed
number of degrees of freedom for interacting gauge fields once SLIV occurs.
Note that a superfluous restriction put on vector or tensor fields would
make it impossible to set the required initial conditions in the appropriate
Cauchy problem and, in quantum theory, to choose self-consistent equal-time
commutation relations \cite{ogi3}. One could expect, however, that quantum
gravity could in general hinder the setting of the required initial
conditions at extra-small distances. Eventually, this would manifest itself
in violation of the above local invariance in a theory through some
high-order operators stemming from the quantum gravity influenced area,
which could lead to physical Lorentz violation. This attractive point seems
to deserve further consideration.

\section{Acknowledgments}

We would like to thank Colin Froggatt, Oleg Kancheli, Archil Kobakhidze,
Rabi Mohapatra and Holger Nielsen for useful discussions and comments.
Financial support from the Georgian National Science Foundation (grant \#
07\_462\_4-270) is gratefully acknowledged by J.L.C. and J.G.J.

\end{document}